# PREDICTIVE GREEN WIRELESS ACCESS: EXPLOITING MOBILITY AND APPLICATION INFORMATION

HATEM ABOU-ZEID AND HOSSAM S. HASSANEIN, QUEEN'S UNIVERSITY

## ABSTRACT

The ever increasing mobile data traffic and dense deployment of wireless networks have made energy efficient radio access imperative. As networks are designed to satisfy peak user demands, radio access energy can be reduced in a number of ways at times of lower demand. This includes putting base stations (BSs) to intermittent short sleep modes during low load, as well as adaptively powering down select BSs completely where demand is low for prolonged time periods. In order to fully exploit such energy conserving mechanisms, networks should be aware of the user temporal and spatial traffic demands. To this end, this article investigates the potential of utilizing predictions of user location and application information as a means to energy saving. We discuss the development of a predictive green wireless access (PreGWA) framework and identify its key functional entities and their interaction. To demonstrate the potential energy savings we then provide a case study on stored video streaming and illustrate how exploiting predictions can minimize BS resource consumption within a single cell, and across a network of cells. Finally, to emphasize the practical potential of PreGWA, we present a distributed heuristic that reduces resource consumption significantly without requiring considerable information or signaling overhead.

## INTRODUCTION

The unprecedented growth of mobile traffic has made energy efficiency in wireless networks of paramount importance. Moreover, the increasing size of mobile devices and variety of tablets, pads, and ultrabooks, is adding more pressure to the energy drain at both the network and end user. Consequently, research and standardization efforts are focusing on devising green mechanisms to save energy across all network elements, as well as user devices [1].

Among the wireless network elements, Base Stations (BSs) account for more than 50 percent of the network energy consumption [2]. Therefore, devising green radio access strategies is vital for overall energy savings. Such mechanisms include: 1) time domain solutions that put BSs to intermittent low energy operating modes during times of inactivity, 2) frequency domain approaches that minimize the operating bandwidth, and 3) network reconfiguration techniques that put BSs to deep sleep and reduce the number of active BSs during low load [2]. Evidently, if the network is aware of the temporal and spatial user traffic demand, it can make better adaptations that reduce energy consumption. The main objective of this article is to present the functional and operational requirements of a predictive green wireless access (PreGWA) framework that exploits user mobility trajectories and application information. The basic idea is that if a user's future location is known, the upcoming data rates can be anticipated from coverage maps, and then used to make energy-efficient rate allocation and network configuration plans.

A key motivation to investigate the use of such predictions for green wireless access is the plethora of information available in smart phones and the evolution of self-organization in networks. The idea is to collect and exchange user location and application information and then develop cooperative access strategies that save energy without compromising user satisfaction. While such predictions were found to provide promising Quality of Service (QoS) gains at high network load [3], the following use cases illustrate their potential for energy-efficient wireless access:

- A user running a delay-tolerant application and moving *towards* a BS can be delayed transmission until getting closer. This allows the BS to save energy by 'sleeping' as the user approaches, and then 'waking up' for a *short* period, during which a *high* data transmission is possible.
- A user viewing a stored video (e.g. YouTube, Netflix) on a highway traversing two cells, can be pre-allocated the requested video content in the first cell, while the second cell is switched off without causing any video stalling.
- A user handed over to a target cell, can report the statuses of the running applications, (e.g. the amount of stored video) and user speed. If the buffer is not empty and speed is considerable, transmission can be momentarily suspended until the user approaches the cell center, without degrading the QoS.

Clearly, the derived gains of PreGWA are dependent on the knowledge of the application's requirements, the user's mobility trajectory, in addition to cooperation between BSs and users.

The organization of this article is as follows. First, we review the fundamental traffic-aware wireless access mechanisms for energy reduction. Next, we present the PreGWA framework and

This work was made possible by a National Priorities Research Program (NPRP) grant from the Qatar National Research Fund (Member of Qatar Foundation).

outline the functional entities required for its development. Then, to demonstrate the potential of PreGWA we consider a case study on energy-efficient delivery of stored video streaming, which is formulated as an optimization problem. Finally, a distributed heuristic that requires minimal predictions and signaling is then presented to illustrate the potential gains of PreGWA in a practical setting.

## TRAFFIC AWARE ENERGY EFFICIENT RADIO ACCESS

Radio access power consumption is distributed among the BS power amplifier (PA), signal processing, air conditioning, and the power supply [2]. Power consumption can be reduced in a number of ways including improving individual hardware component efficiency and developing more efficient baseband signal processing schemes. However, an alternative approach is to power down the hardware components themselves during low traffic load. This can be referred to as traffic aware energy efficiency, as it requires knowledge of traffic demand such that user service is not compromised [4]. The importance of such techniques is that networks are generally designed to support peak demands, which last for only a small fraction of the day. During the rest of the operation time, when traffic demand is lower, BS power consumption can be reduced in a number of ways [2]:

***Time Domain Approaches –*** BSs transmit data to users by allocating units of bandwidth over time slots. During low to medium traffic, the number of required units decreases. Unused resources can be aggregated in a way that creates complete time slots without any data transmission, enabling transceiver hardware to be deactivated to save energy during micro-sleeps of up to 214 μsec. Additionally, more advance sleep modes that reduce the transmission frequency of reference and control signals have been proposed, such as the extended-cell discontinuous transmission (DTX) in Long Term Evolution (LTE). The benefit of such sleep modes is dependent on how fast deactivation and reactivation can be supported by the PA, power supply, and signal processing [2].

Similarly, the user equipment (UE) can also enter a discontinuous reception (DRX) mode to save power by monitoring the downlink (DL) control channel less frequently, and going to sleep when there is no packet scheduled for the UE. A recent comparative study on 3GPP UE sleep mechanisms is available in [5].

***Frequency Domain Approaches –*** As BS transmit power is distributed across the bandwidth, scheduling only a limited number of subcarriers at a given time allows the BS to lower the PA supply voltage, resulting in energy savings. However, the PA is not completely shut down, and therefore energy savings are limited [2]. However, BSs can be implemented such that groups of carriers are served by individual PAs. In this case, the unused PAs can be turned off completely when the corresponding aggregate carriers are not scheduled for transmission. This is known as the carrier aggregation approach in LTE [2].

***Network Reconfiguration –*** Although the aforementioned approaches provide energy savings, BSs still consume a considerable amount of fixed load-independent power to remain functional. Therefore, significant energy savings can be obtained if a select subset of BSs is dynamically switched off completely, when traffic is low for an extended period of time [6]. However, services will be affected during this setup, and a transitional period is needed before inactive BSs can re-operate. Nevertheless, this operation is particularly suited for heterogeneous network deployments, where a macro cell overlays smaller cells that serve users during high traffic demand. When load is low, it is possible to selectively switch off some of the smaller cells, while radio coverage is guaranteed by the macro-cell.

## POTENTIALS OF COOPERATIVE MECHANISMS

Network cooperation is envisioned to play an eminent role to improve network efficiency and long-term user experience [7],[8]. Furthermore, self-organizing networks (SON) introduced in the 3GPP TS 32.521 [9] enable heterogeneous mobile networks to self-optimize and reconfigure, thereby improving user experience and reducing network operational and management costs. Studies have shown that such cooperation can further improve energy efficiency by increasing the potential savings of traffic-aware radio access. For example, in [10] inter-BS cooperation is proposed to optimally determine on-off switching, while in [11] cooperation is used to dynamically adjust cell sizes according to traffic load.

## PREDICTIVE GREEN WIRELESS ACCESS

The proposed PreGWA framework is based on three operational stages 1) collecting and exchanging user location and application information, 2) predicting long-term user traffic demands, and 3) developing cooperative access schemes that save energy without compromising user satisfaction. Figure 1 summarizes the typical sequence of events and information exchanges required in PreGWA, while Fig. 2 outlines the necessary functional entities and their interaction which we discuss below.

***User Location Prediction –*** LTE networks support a range of location positioning methods with varying levels of granularity including the assisted-Global Navigation Satellite System (A-GNSS). A dedicated LTE Positioning Protocol (LPP) is also devised to coordinate signaling between the UE and the BS [12].

The *prediction* of user trajectories is then possible by mapping position, speed and direction of travel onto street maps, particularly for highways and rural areas. To predict mobility patterns for a longer time interval, user input of the destination may be provided either directly or through navigation software. Generating databases of user profiles also facilitates such predictions as recent behavioral studies have shown that people are inclined to follow particular routes with a predictability of up to 93 percent [13]. Moreover, for users on public transportation, the routes of buses and trains are known in advance. As shown in Fig. 2, to provide more accurate predictions, input from real-time traffic information can account for variations arising from the time of day, weather, or accidents. The derived user trajectory predictions can then be used to estimate the spatio-temporal load variability in a network and identify locations where BSs can be switched off.

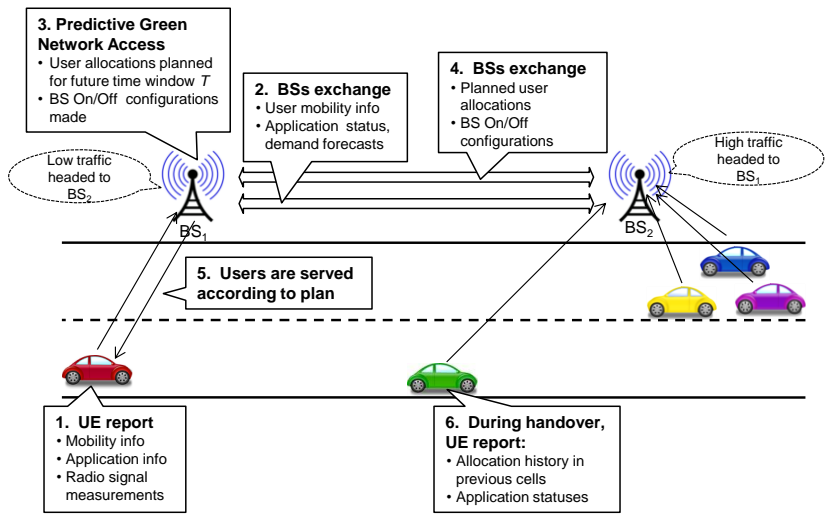

**Figure 1.** *Operations and information exchange in predictive green wireless access.*

***Rate Prediction via Radio Maps –*** While location predictions are useful for BS On/Off configurations, predicting the future rates users will experience can further optimize efficient transmission. This is facilitated by *radio maps* that store average values of historic signal strengths at different geographic locations. Such maps are traditionally generated by the network operator based on road drive tests. A recent feature known as "Minimization of Drive Tests (MDT)" is defined in LTE Rel-10 [14] that exploits the ability of the UE to include location information as part of the radio measurement reporting. The UE logs radio measurements during idle states and sends periodic reports. Additionally, openly accessible radio maps are being crowd sourced from UEs and are available online such as the OpenSignal Project[1].

Coupling radio maps with user location predictions enables the estimation of the average data rates users will experience along a trip. Therefore, the objective of the rate predictor in Fig. 2 is to generate a vector of predicted rates that a user will experience based on the corresponding predicted trajectory.

***Application Demand Information –*** In this module, future rate requirements of users are projected and the status of running applications is classified. For example, the long-term user demand of a stored video stream is predictable based on the streaming rate and duration of the requested video. While the network can infer QoS needs based on the type of traffic, the UE can provide additional information such as application background/foreground running status, as well as user preferences such as quality vs. delay for adaptive streaming videos. In this context, user application profiling can aid the network by registering user preferences and habits to provide additional input to the demand predictor.

As illustrated in Figs. 1 and 2, user application needs are exchanged between BSs to make long-term allocations and network configuration plans. Further, during hand-over, UEs can update the target cell with the application statuses and allocation history in previous cells, which is particularly useful for distributed operation that minimizes BSs communication.

***User Signaling –*** User signaling plays a central role in the PreGWA as shown in Fig. 2. UEs can be used to aid each of the aforementioned prediction processes, i.e. trajectory prediction, rate prediction and application demand forecasting. Provisions for UE involvement in energy efficiency are already being adopted in 3 GPP Release 11. For example, in DRX, the UE assists in determining the favorable connection states. This is because it has the relevant information on the applications running and remaining battery power of the device [5]. However, protocols will be needed in PreGWA to efficiently communicate the multitude of UE information and context to BSs.

***Inter-BS Cooperation –*** BS cooperation is required at two levels. The first is for the exchange of the UE gathered information at each site, and the second is to collaborate in making the green access allocation decisions and network configuration. The amount of communication depends on whether access decisions are made centrally or are distributed, and on the presence of any iterative procedures to converge to a final decision. Such cooperation can be facilitated in LTE over the special inter-BS X2 interface that allows some form of communication and coordination between BSs.

***Green Network Access Engine –*** As depicted in Fig. 2, the generated predictions are forwarded to the green network access engine to devise energy efficient transmission and network configurations. Also note that the network configuration output of the green network access engine is fed back to the operator radio map to account for the changes in the network layout. In the access engine, the resource sharing model is first defined, e.g. a time slotted system where users can share air-time in arbitrary fractions in each slot. Then, depending on whether time or frequency domain resource sharing is implemented (or both), the load-power consumption model is devised as in [15]. To incorporate BS On/Off switching, the value of deep sleep power consumption is required, in addition to constraints on the minimum time duration needed before a BS can be turned back on. Thereafter, based on the user application requirements and corresponding rate predictions, the access engine determines 1) how BS resources will be allocated among users over a predefined prediction window *T*, and 2) the BS On/Off configuration for each BS during *T*. The objective is to minimize power consumption without compromising user QoS.

---

[1]. http://www.opensignal.com

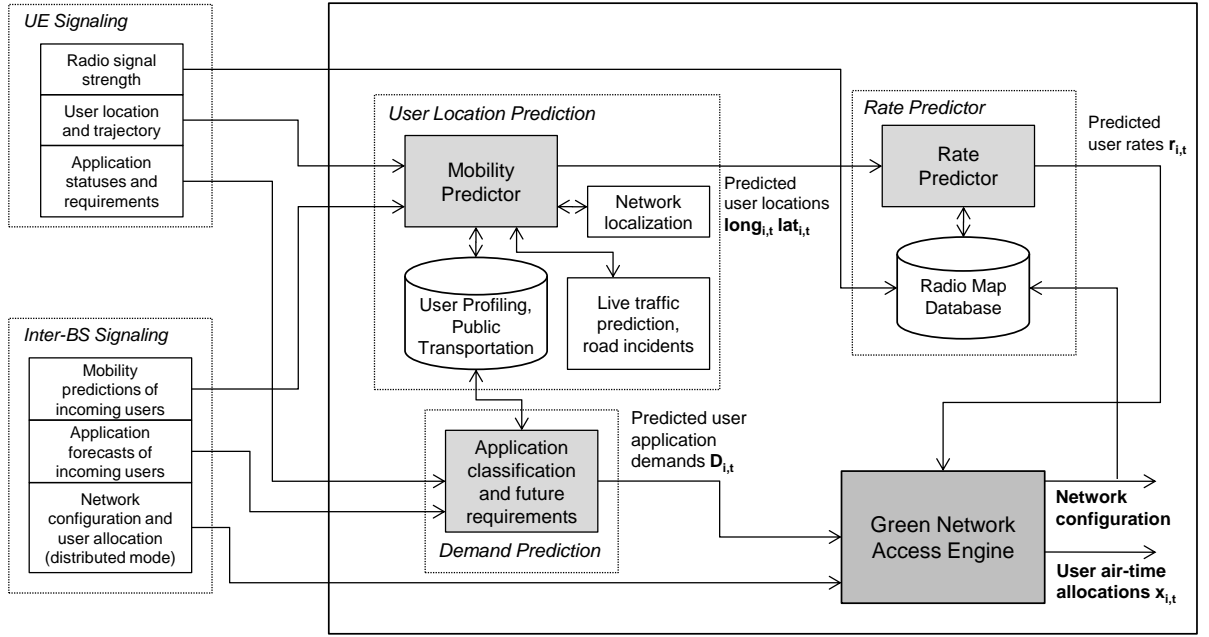

**Figure 2**. *Key elements and functions of PreGWA. The subscripts i and t represent users and time instances respectively.*

Figure 3 provides a conceptual output of the allocation engine for a time slotted system where the air-time fraction $x_{i,t}$ for each user, is defined over *T* such that the sum air-time over all users is minimized. As PreGWA covers multiple cells, there are two modes of operation as shown in the figure. When operating in a centralized mode, a coordinating BS makes the allocation plans for all the collaborating BSs, whereas in distributed operation each BS makes its own allocation plan but information can be exchanged between BSs during this process.

To provide a better understanding on formulating a green network access scheme and the potential energy savings, we present a case study on stored video streaming in the following section.

# A CASE STUDY: STORED VIDEO STREAMING

Recent forecasts on mobile video project a 16-fold growth from 2012 to 2017, with video accounting for 66 percent of global mobile traffic[1]. Moreover, the growing usage of larger screen devices, the development of in-vehicle infotainment systems, and the increasing transit commute times, motivate the development of green access schemes for vehicular video delivery.

In stored video streams such as HTTP based streaming, the requested content is delivered as data chunks containing the video segments that are stored momentarily before being played back at the UE. As opposed to live streaming, the content can be delivered ahead of time and stored at the UE, after which transmission can be momentarily suspended while the user consumes the buffer. By jointly utilizing the user location and rate predictions over multiple cells, network energy consumption can be minimized as discussed below.

## SYSTEM MODEL

We consider the simple scenario of two cooperating BSs covering the road shown in Fig. 1 with one way road traffic from $BS_1$ to $BS_2$. Let *M* denote the BS set $M = \{1,2,..., M\}$, and *N* denote the active user set $N = \{1,2,..., N\}$. For simplicity, users are associated to BSs based on the closest distance. The set $U_{j,t}$ contains the indices of all the users associated with BS *j* at time *t*. To provide realistic mobility, vehicle traces are generated using the SUMO microscopic traffic simulator[2]. Average user received power is computed based on the path loss model PL(d) = 128.1 + 37.6 log10 d, where d is the user-BS distance in km. As we focus on formulating the green video transmission scheme, the user-BS distance is assumed to be known during the window of *T* seconds. Time is divided in equal slots of duration *τ*, during which the path loss is assumed to be constant. A typical value of such a coherence time *τ* is 1 s for vehicle speeds up to 20m/s, during which path loss is not significantly affected. The feasible link rate is computed using Shannon's equation with SNR clipping at 20 dB for practical modulation orders. Therefore, a user *i* at slot *t*, will have a feasible data transmission of:

$$r_{i,t} = \tau B \log_2(1 + P_{rx}/N_o B) \qquad (1)$$

where $P_{rx}$ and $N_o$ are the received power and noise power spectral density respectively, and *B* is the bandwidth. In this illustrative example, we assume that the predictions of $r_{i,t}$ are accurate, to demonstrate the potential gains in the error free scenario.

***Stored Video Traffic Model –*** A user's traffic demand for a stored video can be represented by the cumulative video content required for smooth

[1.] *Cisco Visual Networking Index: Global Mobile Data Traffic Forecast Update, 2012-2017.*

[2.] *Simulation of Urban Mobility (http://sumo.sourceforge.net)*

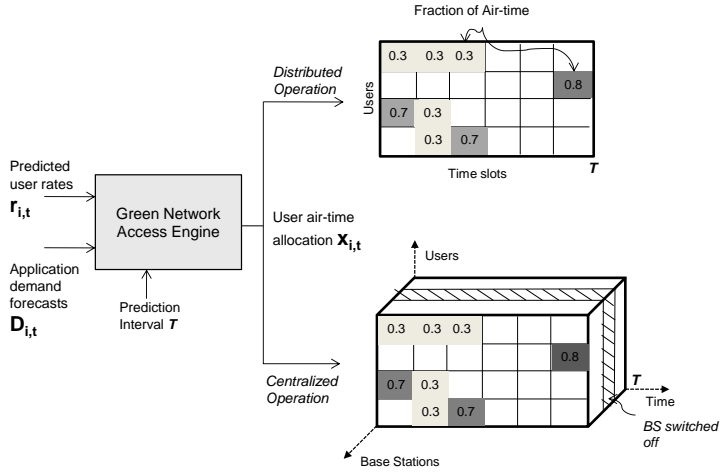

**Figure 3.** *User air-time planning in PreGWA. In distributed operation, each BS makes its own allocation plan; whereas in centralized operation, a coordinating BS makes the allocation plans for all the collaborating BSs.*

playback. For example, if a user requests a video with a streaming rate $V$ [bps] at a time $t=0$, then at any time $t_x$, the delivered cumulative content should not be less than $Vt_x$ for continuous playback. This is illustrated in Fig. 4 by $D_{i,t}$. Note that it is possible to transmit more than $Vt_x$, which is buffered at the UE.

**Resource Sharing Factor** – During $\tau$, the BS resource elements can be shared in arbitrary ratios between the users. We define the resource sharing factor $x_{i,t} \in [0, 1]$ as the fraction of time that the BS resources are assigned to user $i$ at each time slot $t$. The received data rate is then $x_{i,t} r_{i,t}$. Lower air-time corresponds to less energy consumption as the BS will have more sleep opportunities.

## OPTIMAL PREDICTIVE GREEN ACCESS

In Fig. 4a we illustrate how a conventional allocation scheme that distributes BS air-time equally among users will continue to serve users in poor channel conditions. Notice how the user cumulative rate allocation denoted by $R_{i,t}$ is significantly higher than the required allocation $D_{i,t}$ implying that the video content is being pre-buffered even when channel rates are poor. On the other hand, as shown in Fig. 4b, a scheme aware of the user's future rate, will plan to prebuffer as much content as possible at times of high channel conditions, while ensuring that the cumulative content requested is not violated when the user is in poor channel conditions. This achieves a much lower air-time usage. Figure 4c, demonstrates that it is also possible to consider the case where $BS_2$ is switched off, by modifying the predicted rate to account for this. In this case, it is possible to completely pre-buffer the video content before the user leaves $BS_1$. Although the total air-time is higher than that in Fig. 4b, this can achieve more energy savings as $BS_2$ is switched off completely.

In the following formulation we consider the multi-user, multi-cell scenario, with a central coordinating BS which plans air-time allocations for all the cooperating BSs. The problem of minimizing network air-time over $T$, which we refer to as Air-time Min-Optimal, can be expressed as the following Linear Program (LP):

$$Minimize: \sum_t \sum_i x_{i,t}$$

$$Subject\ to: \sum_{i \in U_{j,t}} x_{i,t} \leq 1 \quad \forall j, t$$

$$D_{i,t} - R_{i,t} \leq 0 \quad \forall i, t$$

$$\sum_t x_{i,t} r_{i,t} \leq D_i \quad \forall i$$

$$0 \leq x_{i,t} \leq 1 \quad \forall i, t. \quad (2)$$

The first constraint captures the BS resource limitation and the second constraint specifies the cumulative rate requirement of each user. When $D_{i,t} < R_{i,t}$ it implies that the user has future video content buffered, and it is this buffering capability that facilitates making the optimal allocations that minimize air-time. The third constraint limits the amount of video content assigned to each user, to the total size of the requested video $D_i$, and the last constraint defines the bounds for the resource sharing factor.

Notice that this optimization problem is coupled over multiple BSs. This is due to 1) the user air-time is computed over multiple BSs as the objective function minimizes the *network* air-time, and 2) the cumulative user rates computed in the second constraint require BSs to be aware of the allocations made to users in previous BSs. Therefore, the allocated content to users in one BS, impacts the future amounts allocated in upcoming BSs that the user will traverse.

As an extension to the formulation in Eq. 2 we can incorporate switching BSs completely off as well. In this case a trade-off between minimizing the total network time and minimizing the total number of active BSs needs to be included in the objective function.

Although the formulation in Eq. 2 is an LP, generating the constraint matrix has a very large memory requirement and requires significant computational power due to the long-term planning horizon and multiple cells involved. Further, as it is centralized, a signaling overheard is incurred. We therefore present the following lightweight, distributed heuristic that achieves close to optimal performance at low load.

## DISTRIBUTED HEURISTIC

As previously illustrated in Fig. 4, BS air-time is minimized when users are granted more air-time access at their highest achievable data rates and less access when they are at lower achievable rates. This allows the BS to transmit more data in less time. The following distributed heuristic is divided into two stages. In the first stage, minimum air-time is granted to each user, to ensure smooth playback (i.e. $D_{i,t} = R_{i,t}$). If $R_{i,t-1} > D_{i,t}$ then no air-time is granted to this user in this stage. In the second stage, excess BS air-time is allocated to users whose achievable data rate is going to decrease (i.e. they are moving away from the BS). This is to opportunistically pre-buffer as much video content as possible before the user's achievable rate decreases any further. The heuristic, which we refer to as Air-

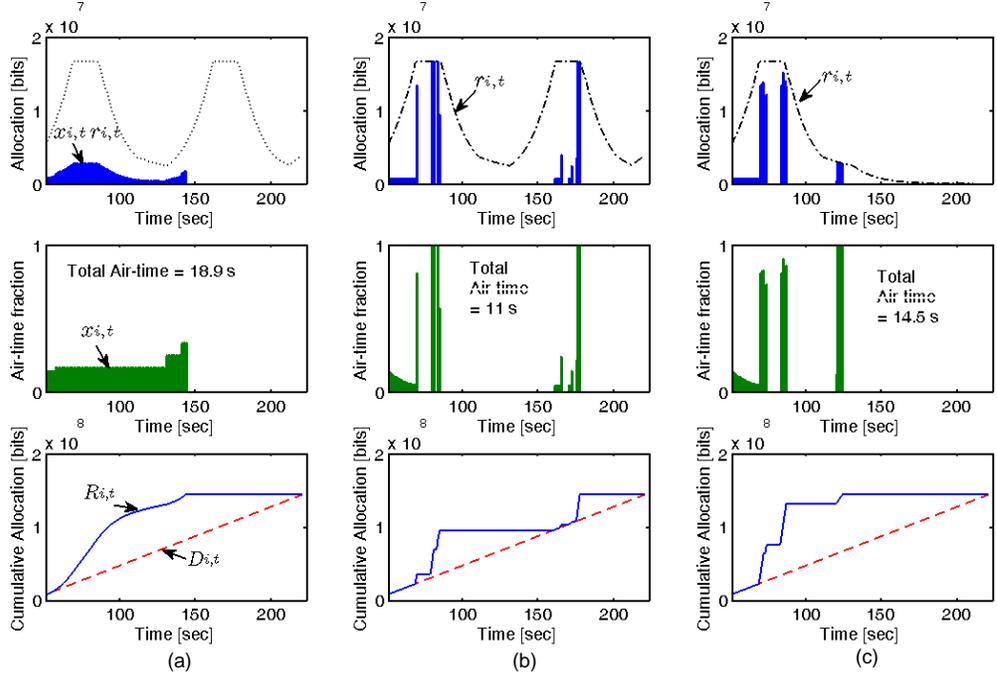

**Figure 4.** *Illustration of air-time minimization for multi-user stored video streaming: (a) conventional allocation that divides air-time equally without exploiting rate predictions and video buffer status; (b) air-time minimization using predictions: allocations are low if the user rate is increasing, and high when the user rate at a peak to pre-buffer content efficiently; (c) similar to (b) but using only a single BS to provide all the video content in $BS_1$ Although air-time in (c) is larger than (b), more energy is saved as one BS is switched completely off.*

time Min-Heuristic performs the following steps at each BS in every time slot:
- **Step 1:** Sort the users in descending order of their achievable rates.
- **Step 2:** Grant each user the required air-time that satisfies constraint 2 in Eq. 2. This is computed as: $x_{i,t} = \max(0, D_{i,t} - R_{i,t-1})/r_{i,t}$, where $r_{i,t}$ is the current achievable rate.
- **Step 3:** Determine the subset of users, whose rate is going to decrease, and that have remaining video content to be delivered. If the set is empty, end the step; otherwise sort the set in descending order of the achievable user rates.
- **Step 4:** Use the remaining BS air-time to pre-buffer as much future video content as possible, to the first user in the sorted set of step 3. Remove this user from the set of users with decreasing rates.
- **Step 5:** Repeat step 4 if there is additional BS air-time.

### Remarks:
*1.* In step 2, generally speaking, it not necessary to allocate BS air-time sequentially to the ordered users, as there should be sufficient air-time to serve all the users with their minimal video content (as we are considering saving air-time during *low* network load).
*2.* Also in step 2, during handover when a user changes BS association, some signaling is required to enable the target BS to compute $x_{i,t} = \max(0, D_{i,t} - R_{i,t-1})/r_{i,t}$. This is because $R_{i,t-1}$ is unknown to the target BS, and should be signaled either from the UE or from the source BS. Additionally, note that $x_{i,t}$ can be computed as $\max(0, V - Buff_{i,t-1})/r_{i,t}$ where $Buff_{i,t-1}$ is the amount of video content buffered at the UE. If this is equal to zero, then $x_{i,t} = V/r_{i,t}$, and if it is larger than $V$, then $x_{i,t} = 0$. This implies that the target BS only requires the video buffer status of the incoming user, which is signaled with minimal overhead without any centralized operation.
*3.* No prediction of the entire user rate vector is required. The heuristic only needs to know whether a user rate is increasing or decreasing.
*4.* The heuristic has a very low computational complexity of O($M \log M + M$).

## PERFORMANCE EVALUATION

The potential air-time minimization gains are evaluated for the two cell scenario of Fig. 1. The BSs are 500m from the road edges and 1 km apart. Transmit power is 40 W, over a 5 MHz bandwidth, and $T$ is 200 seconds.

To provide a base-line reference we consider the following allocation schemes that do not incorporate any rate predictions.
**Equal Share (ES):** air-time is shared equally among the users at each time slot.
**Rate-Proportional (RP):** the RP allocator prioritizes users with high rates, while still serving users with poor rates. The user air-time is the ratio of its data-rate to the sum all users rates: $x_{i,t} = r_{i,t}/\sum_{i \in U_{j,t}} r_{i,t}$.

Figure 5a considers the scenario of a large number of users (40), requesting low rate video streams. We can see a very significant reduction of air-time exceeding 70% at low loads. The exact energy savings will depend on the BS power model [15]. As the streaming rate increases, the gains decrease since network utilization increases and the degrees of freedom in allocation decrease. Interestingly, the distributed heuristic achieves near optimal performance, with only minimal user-BS cooperation and rate knowledge. Figure 5a also indicates that if $BS_2$ is switched completely off it is possible to serve the users with $BS_1$ alone at a minor increase in air-time. This corresponds to even greater energy savings as the number of active BSs is reduced.

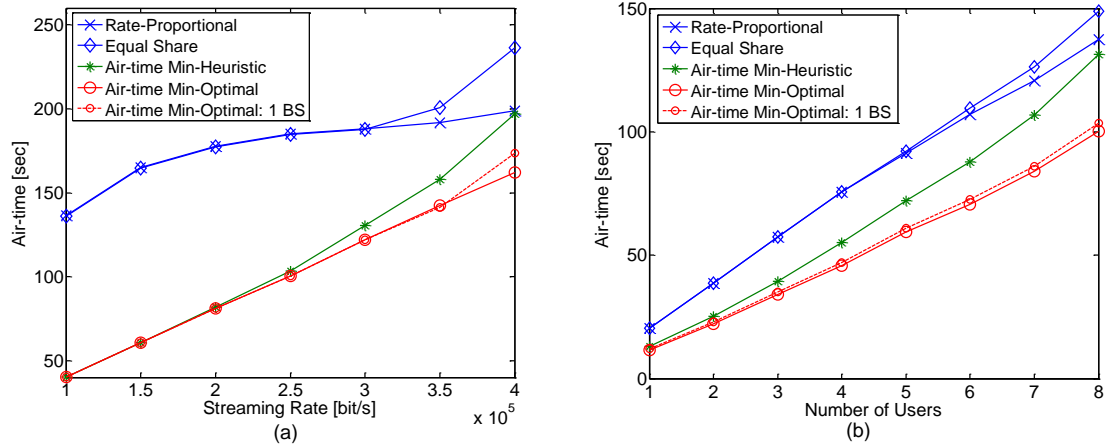

**Figure 5.** *Average air-time with (a) varying streaming rates for 40 users; (b) varying number of users for 1.2 Mbps streaming.*

In Fig. 5b, we investigate the converse case, where a few users are requesting high rate video streams (1.2 Mbps). In this scenario, although the total network traffic is low, it is not possible to pre-buffer significant portions of video content in advance when the achievable rate is high, since the requested rate is also high. Also, when the achievable rate is low, air-time cannot be reduced significantly due to the high user streaming rate. This implies that air-time reduction shall be less in this case, as confirmed in the results. The performance of the distributed heuristic also deviates from the optimal allocation as the network load increases. Nevertheless, it achieves considerable gains at low load, with minimal computation and signaling overhead. However, more advanced heuristics that exploit the rate prediction vector completely are also needed.

## CONCLUSION AND FUTURE DIRECTIONS

In this article, we discussed how user location predictions and application information can be incorporated into the radio access framework to provide energy savings, and outlined the required functional entities to achieve this. As a case study, we applied the predictive green access framework to stored video streaming and formulated the minimum air-time access problem as a linear program. A distributed heuristic was then developed and its performance investigated under varying traffic assumptions. Results indicated that significant energy savings are achievable with minimal prediction requirements and signaling overhead.

There are several directions for future work. First of all, studies on the effects of prediction errors are needed to assess performance in a practical environment. Furthermore, models that capture the variability and accuracy of the predicted information, as well as methods and algorithms that are robust to such inaccuracies are required. Also, resource allocation formulations for applications other than stored video streaming need to be developed, in addition to advanced multi-objective formulations that account for the tradeoff between QoS and energy. Another important issue is that of user-BS and inter-BS signaling, where effective protocols and implementation mechanisms built on existing standard interfaces are needed. Additionally, applying the predictive mechanisms over heterogeneous networks will add to the signaling challenge, and drive the need for close to optimal distributed solutions. Therefore, while we have seen that location predictions and application information can provide valuable wireless access energy savings, several open research issues remain to be investigated to assess the full potential and practical use of such mechanisms.

HATEM ABOU-ZEID (h.abouzeid@queensu.ca) is currently pursuing his Ph.D. degree in Electrical and Computer Engineering at Queen's University, Canada. He received his B.Sc. and M.Sc. degrees (with honors) in communications engineering from the Arab Academy for Science, Technology and Maritime Transport, Egypt, in 2005 and 2008, respectively. From July 2011 to December 2011, he was a research intern at Bell Labs Alcatel-Lucent, Germany. He has served as a reviewer for several international journals and conferences. His current research interests include predictive resource allocation, optimizing video streaming delivery, and energy-efficient system design.

HOSSAM S. HASSANEIN (hossam@cs.queensu.ca) is a leading authority in the areas of broadband, wireless and mobile networks architecture, protocols, control and performance evaluation. His record spans more than 500 publications in journals, conferences and book chapters, in addition to numerous keynotes and plenary talks in flagship venues. Dr. Hassanein has received several recognition and best papers awards at top international conferences. He is also the founder and director of the Telecommunications Research (TR) Lab at Queen's University School of Computing, with extensive international academic and industrial collaborations. He is a senior member of the IEEE, and is currently chair of the IEEE Communication Society Technical Committee on Ad hoc and Sensor Networks (TC AHSN). Dr. Hassanein is an IEEE Communications Society Distinguished Speaker (Distinguished Lecturer 2008-2010).

**FIGURES**

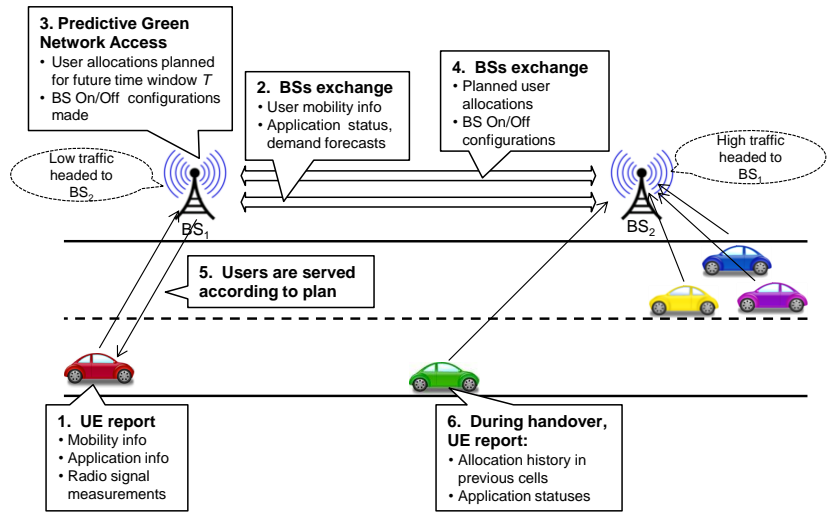

**Figure 1.** *Operations and information exchange in predictive green wireless access.*

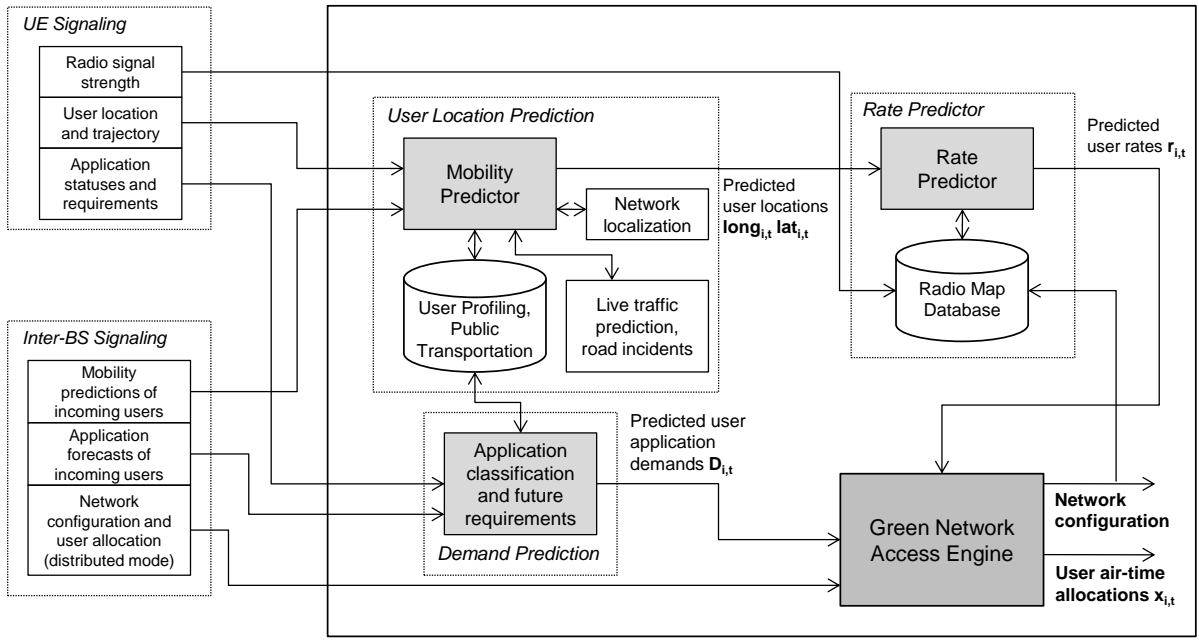

**Figure 2**. *Key elements and functions of PreGWA. The subscripts i and t represent users and time instances respectively.*

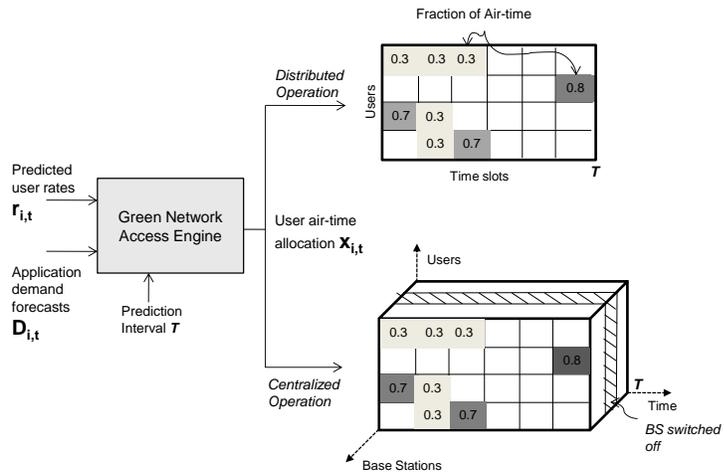

**Figure 3.** *User air-time planning in PreGWA. In distributed operation, each BS makes its own allocation plan; whereas in centralized operation, a coordinating BS makes the allocation plans for all the collaborating BSs.*

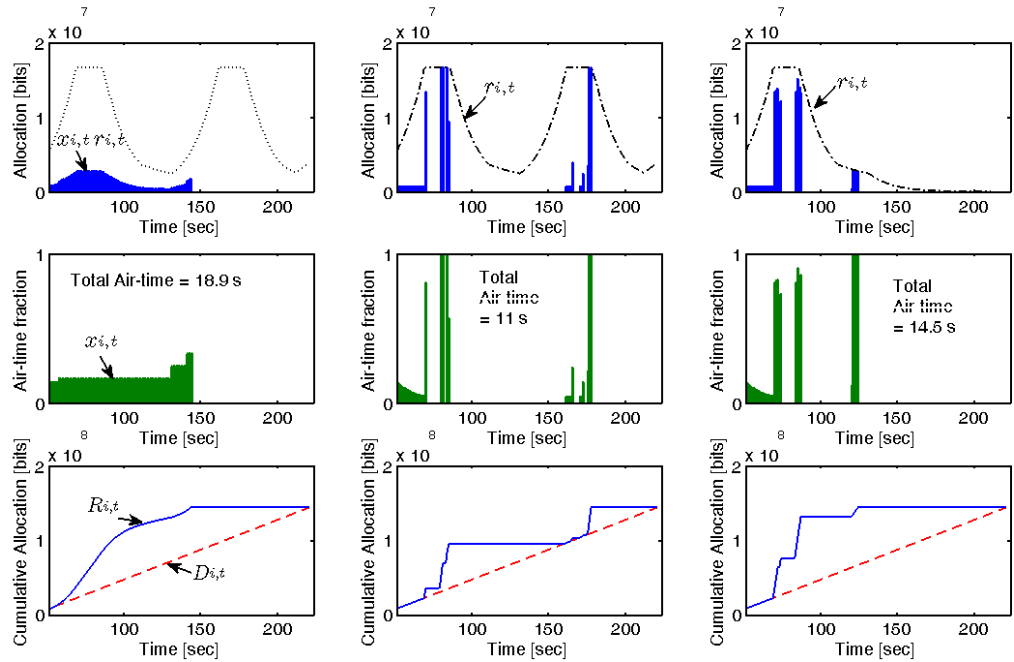

**Figure 4.** *Illustration of air-time minimization for multi-user stored video streaming: (a) conventional allocation that divides air-time equally without exploiting rate predictions and video buffer status; (b) air-time minimization using predictions: allocations are low if the user rate is increasing, and high when the user rate at a peak to pre-buffer content efficiently; (c) similar to (b) but using only a single BS to provide all the video content in $BS_1$ Although air-time in (c) is larger than (b), more energy is saved as one BS is switched completely off.*

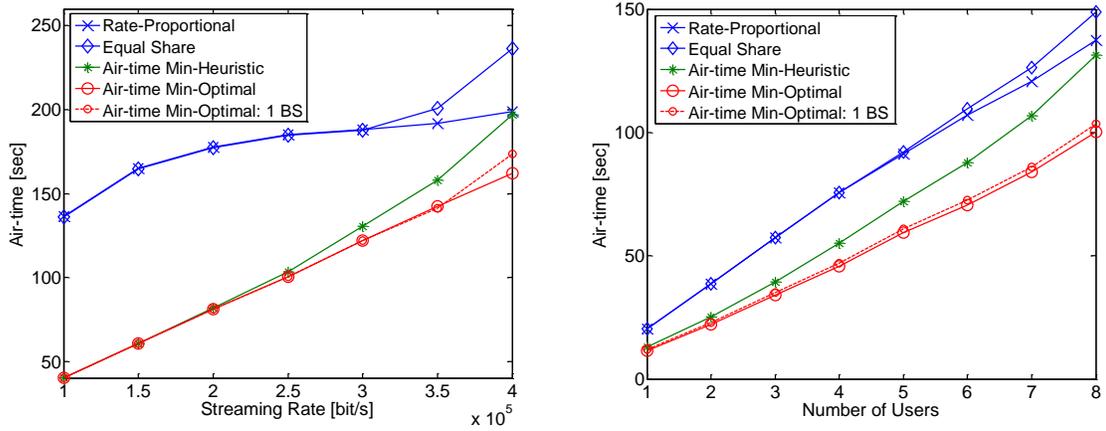

**Figure 5.** *Average air-time with (a) varying streaming rates for 40 users; (b) varying number of users for 1.2 Mbps streaming.*